\documentclass[aps,prl,twocolumn,showpacs]{revtex4}
\usepackage{graphicx}

\usepackage{graphicx}
\begin{document}
\title{Quantum critical phase in BaVS$_3$}

\author{N. Bari\v si\'c$^1$}
\email{barisic@stanford.edu}
\author{I. K\'ezsm\'arki$^2$}
\author{P. Fazekas$^3$}
\author{G. Mih\'aly$^2$}
\author{H. Berger$^1$}
\author{L. Demk\'o$^2$}
\author{L. Forr\'o$^1$}

\affiliation
{$^1$Institut de Physique de la mati\`{e}re complexe,EPFL, CH-1015 Lausanne, Switzerland\\
$^2$ Electron Transport Research Group of the Hungarian
Academy of Science and Department of Physics, Budapest University of Technology and Economics, 1111 Budapest, Hungary\\
$^3$ Research Institute for Solid State Physics and Optics, Budapest 114, P.O.B. 49, H-1525 Hungary}

\date{\today}
\begin{abstract}
We study the high-pressure metallic phase of high-purity single
crystals of BaVS$_3$ by measuring the temperature, pressure, and
magnetic field dependence of the resistivity. Above the critical
pressure of $p_{\rm cr}=1.97$GPa an extended non-Fermi liquid
$p-T$ regime emerges with resistivity exponent $1.5 \le n<2$,
crossing over to a FL only around $p=2.7$GPa. A hysteretic feature
indicates that close to the insulator--metal boundary, the system
is magnetically ordered. Our findings reveal a close analogy
between the extended partially ordered NFL state of
non-conventional itinerant magnets and the corresponding state of
BaVS$_3$.
\end{abstract}


\maketitle

Under atmospheric pressure the linear chain compound BaVS$_3$
undergoes a series of phase transitions \cite{Mihaly}. At
$T_{s}=250$K a hexagonal-to-orthorhombic transition takes place,
at $T_{\rm MI}=70$K a metal-insulator transition (MIT) occurs,
which is accompanied by the doubling of the unit cell along the
Vanadium chains ($c$-axis). Finally there is a magnetic transition
at $T_X=30$K whose main aspect is the onset of an incommensurate
3-dimensional magnetic order with a long-period structure along
the $c$-axis \cite{Inami,Fagot,fagot_structure,TxRef}. The $T<T_X$
phase was observed to break time reversal invariance, while the
other two transitions appear as purely structural
\cite{Inami,fagot_structure}. The MIT is broadly speaking a
commensurate Peierls transition \cite{Emery,Giamarchi} where the
development of a bond/charge order wave \cite{pougetdec} opens a
gap in the Q1d d-band \cite{Mitrovic,Thesis}. Although this
transition has a major influence on the susceptibility, and its
anisotropy, the corresponding magnetic correlations are not
finally settled. Only magnetic quasi-LRO \cite{TxRef} has been
observed so far. If any kind of magnetic order occurs below
$T_{\rm MI}$ the $T_X$-transition is its merger with magnetic LRO
observed at low temperatures \cite{TxRef,Thesis}.

A phase with incipient LRO should be particularly sensitive to a
change of control parameters, including pressure and magnetic
field. In contrast, a phase with well-developed order may be less
sensitive to such changes \cite{Thesis}. This is evident from the pressure
dependence of the phase diagram in Fig 2c: $T_{\rm MI}(p)$ decreases fast
with increasing pressure, and the insulating phase is vanishing at
a critical pressure $p_{\rm cr}\approx 2$GPa \cite{Forro}. The
transition temperature of the magnetic ordering, $T_X(p)$, is
$p$-independent at least up to $\sim 0.7$GPa
\cite{nakamura_private}, admitting the possibility that the MIT
and the magnetic transition merge slightly below 2GPa. Indeed,
our recent high-pressure magnetoresistivity studies indicated that
in the vicinity of the critical pressure the structural transition
and the magnetic ordering phenomenon combine, and in presence of
magnetic field a complex order is sought by a hierarchy of
very slow relaxation processes \cite{Hysteresis}.

When the insulating phase is completely suppressed by pressure
signs of non-Fermi liquid (NFL) behavior emerge \cite{Forro},
similar to those observed in heavy fermion systems near to quantum
critical points (QCPs). A quantum phase transition may arise
whenever the critical temperature $T_{\rm cr}$ of some kind of
long range order is suppressed by a control parameter. In heavy
fermion systems, the suppressed order usually has
antiferromagnetic character, and $T_{\rm cr}\to 0$ gradually as
$p\to p_{\rm cr}$ within the metallic phase. The NFL regime is
observed at $p_{\rm cr}$ in a limited part of the parameter space;
the Fermi liquid behavior returns as the pressure is further
increased from $p_{\rm cr}$. The canonical picture of NFL behavior
resulting from a nearby QCP would ascribe it to a wedge-shaped
region in the $T-p$ plane, in particular, also to $T\to 0$ at
$p=p_{\rm cr}$ \cite{ueda_moriya}. The relationship to underlying
microscopic models is complicated \cite{Thesis} but let us note
that BaVS$_3$ shares a basic feature with several other NFL
systems: the coexistence of wide-band and narrow-band states near
the Fermi level \cite{Mitrovic,Lechermann}.

In this paper we investigate the non-Fermi liquid behavior in
BaVS$_3$ by varying the control parameters p and B. We show that
in this compound the suppression of the combined magnetic and
structural order in a quantum phase transition is unlike the
simpler QCP phenomena studied previously in heavy fermion systems.
Our results reveal NFL behavior not only at the critical pressure,
but rather in an extended range of pressure above $p_{\rm cr}$, at
sufficiently low temperatures and magnetic fields. The large area
covered by the NFL state implies that it does not result from a
relatively distant QCP, but rather it is a critical phase as
suggested for certain nearly/weakly ferromagnetic systems
\cite{nonFL_FM,FLbreakdown_MnSi}. Some relevant features of this
critical phase is also revealed by high-field magnetotransport
measurements.

\begin{figure}[t!]
\centerline{\includegraphics[width=0.9\columnwidth] {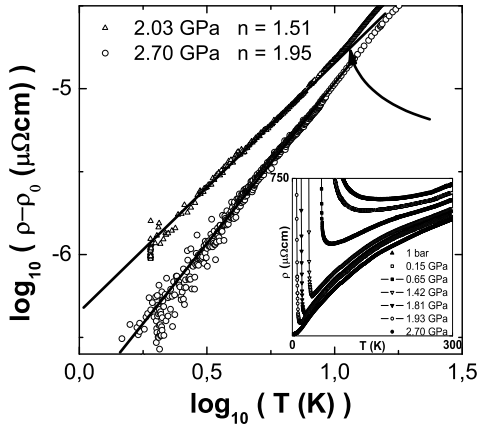}}
\caption{The low temperature (up to 40 K) part of the resistivity
at 2.03 GPa, presented in the log-log plot, reveals the power-law
dependence below 15 K. Position of the shoulder (~15 K) is
indicated by the arrow. The power-law fits (indicated by the
straight lines) give the $T^n$ coefficients $n=1.5$ and $n=2$ at
pressures of 2.03 and 2.7 GPa respectively. Inset: Temperature
dependence of resistivity for various pressures showing the
suppression of $T_{MI}$}. \label{fig1}

\end{figure}

In an effort to accurately determine the intrinsic low energy
properties of BaVS$_3$, special care was taken to use high-quality
single crystals. After obtaining crystals by the previously
established Tellurium flux method \cite{Kuriyaki}, a careful
characterization was carried out. The single crystals were
selected on the basis of the following criteria: i) resistivity
measurements at ambient pressure were required to exhibit metallic
behavior at high temperatures with a well-defined change of the
slope at $T_S$, a sharp MI transition and no sign of saturation of
resistivity in the insulating phase; ii) the resistivity measured
at 2 GPa (i.e. at the lowest pressure where the metallic phase
extends to whole temperature range) needed to display a high
residual resistivity ratio $\rho_{\rm RT}/\rho_{T=2{\rm K}}\sim
50$; iii) magnetic susceptibility anisotropy was required to
exhibit both low temperature transitions ($T_{\rm MI}$ and $T_X$
clearly and sharply and no Curie tail at temperatures below 10 K.
Samples with a typical dimensions of 2 x 0.2 x 0.2 mm$^3$ were
mounted into a nonmagnetic self-clamped pressure cell where
kerosene was used for the pressure transmitting medium. The pressure
was monitored in situ using a calibrated InSb pressure gauge. This
experimental setup enables resistivity and magnetoresistance
measurements over a wide range of temperatures (1.5 to 300 K),
pressures (up to 3 GPa) and magnetic fields (up to 12.7 T). The
resistivity is measured along the chain direction and the applied
magnetic field is perpendicular to the chains. The sweeping rate
of the magnetic field was kept slow enough to avoid the heating
due to eddy currents.

Figure 1 shows the temperature dependence of the resistivity
measured at various pressures. In accordance with previous studies
\cite{Forro} the MIT is suppressed at $\approx 2$GPa. In the
following, we investigate exclusively the high-pressure metallic
phase above the critical pressure of $p_{\rm cr}=2 \pm 0.02$GPa,
where the low temperature resistivity follows a fractional power
law behavior with exponent $1.5\le n \le 2$. This customary
signature of non-Fermi liquid behavior extends over a broad
temperature range (up to 15 K), and a wide pressure interval. Figure 1
displays the two limiting cases. At $p=2.03$GPa $n=1.51\pm 0.01$,
the behavior expected of nearly antiferromagnetic systems. At
$p=2.7$GPa $n=1.95\pm 0.05$ indicating that Fermi liquid behavior
is essentially restored.

\begin{figure}[b!]
\centerline{\includegraphics[width=1\columnwidth] {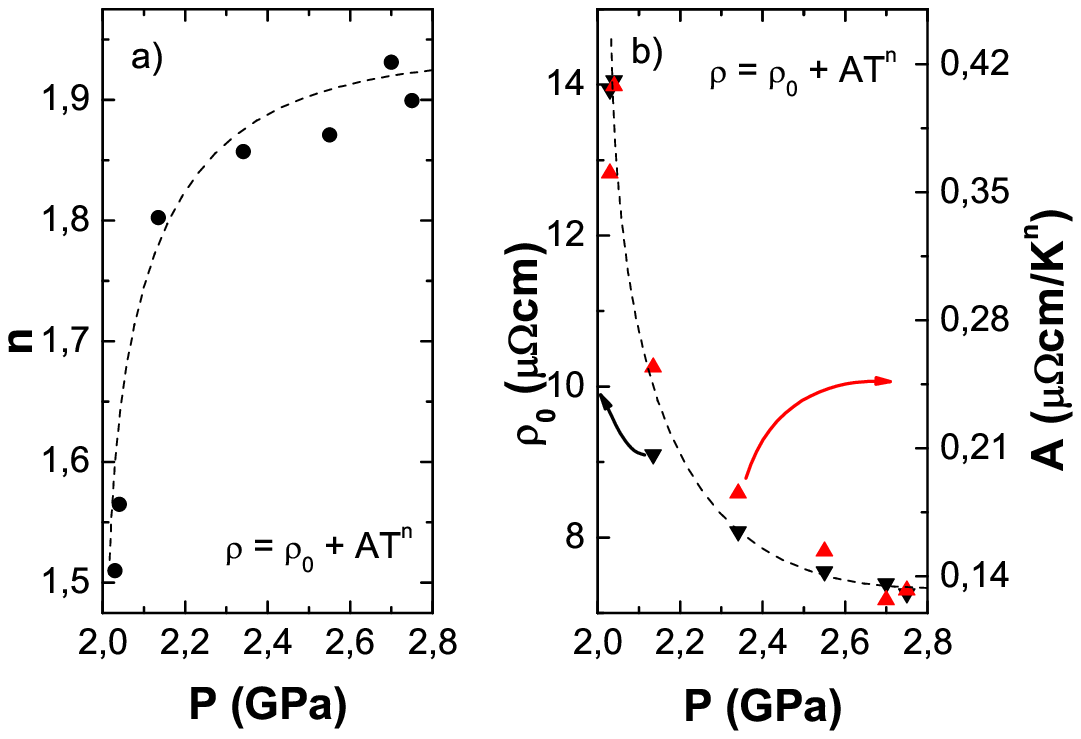}}
\centerline{\includegraphics[width=0.9\columnwidth] {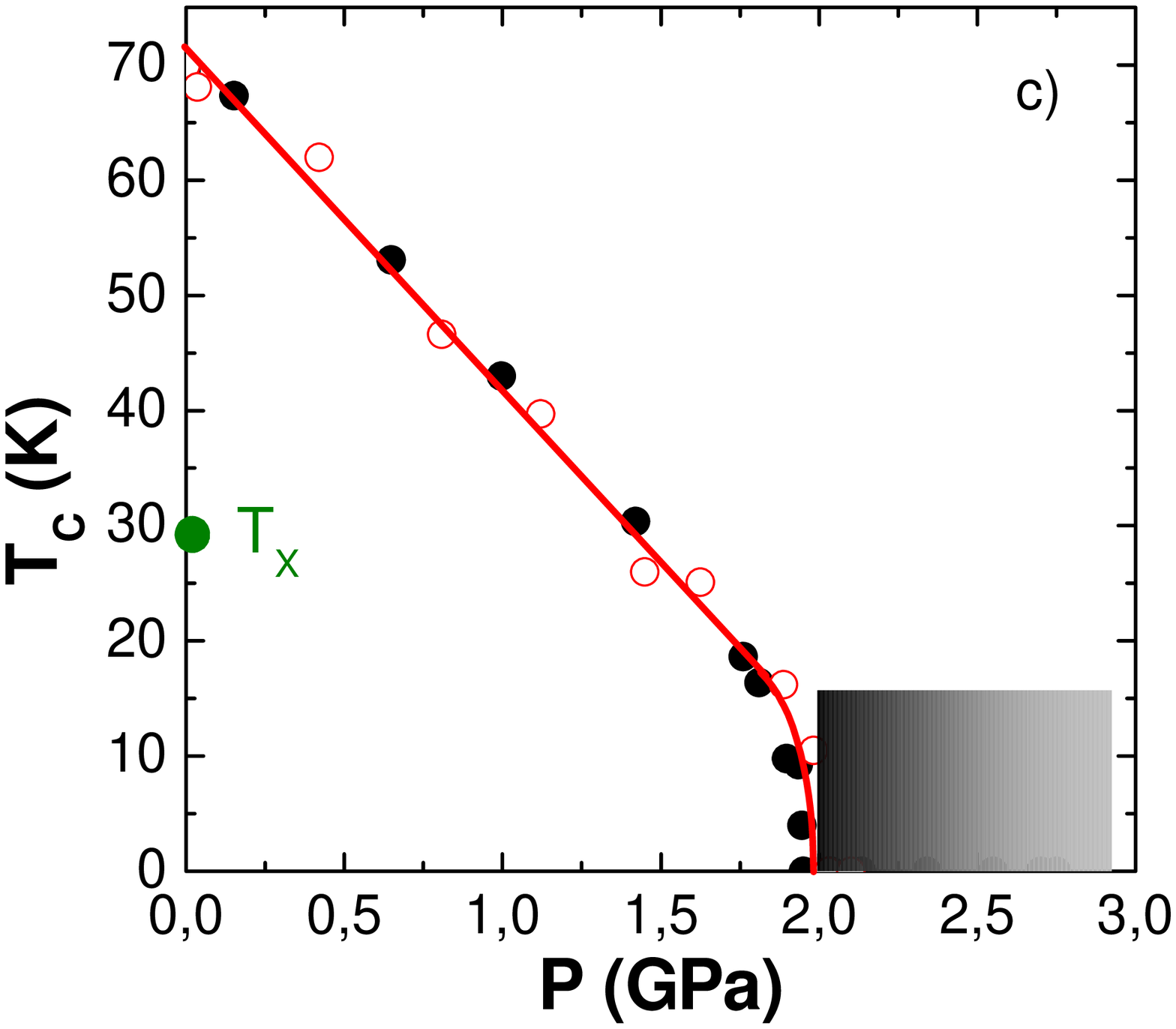}}
\caption{(Color online) Pressure dependence of the a) resistivity
exponent $n$ (the dashed line is to guide the eye), b) Pressure
dependence of the prefactor $A$ (red squares) and the "residual"
resistivity $\rho_0$ (black circles) as the results of the fits
$\rho=\rho_0+AT^n$ in the temperature range $1.7<T<15$ K. c) The
temperature-pressure phase diagram of $BaVS_3$. The MI phase
boundary is marked by the red line. The gray regions are related
to crossover form NFL to FL, the darkest gray symbolizing n =1.5
and the lightest n = 2.}

\label{fig2}
\end{figure}

To map out the entire intervening regime, the low-temperature
resistivity is fitted to
\[
\rho(T,p) = \rho_0(p) + A(p) T^{n(p)}
\]
with $\rho_0$, $A$ and $n$ as free parameters. The results are
shown in Fig. 2. The exponent rises continuously from $n \approx
1.5$ to the Fermi liquid value $n=2$ as $p$ is increased from
$p_{\rm cr}$ to $p= 2.7$Gpa. This power-law behavior with
gradually changing exponent extends over the grey region displayed
in Fig.~2 The pressure dependence of the pre-factor $A$ and the
"residual resistivity" $\rho_0$ is complementary to that of the
exponent and both of them seem to diverge as the pressure
approaches $p_{\rm cr}$.

Plotting the pressure dependence of $A$ and $n$ is standard in the
discussion of NFL systems \cite{Knebel}. However, as shown above,
in order to get reasonable fits one must allow the pressure
dependence of $\rho_0$, as well \cite{Thesis}. The result is shown
in Fig. 2b. Lowering the pressure towards $p_{\rm cr}$, $\rho_0$
doubles to $\rho_0(p=2.03{\rm GPa})=14\mu\Omega{\rm cm}$, i.e.,
the mean free path has been halved to $l\approx 160{\AA}$. Since
extrinsic defects might not appear due to relaxing the pressure,
we ascribe the increment of $\rho_0$ to the interplay of the
anomalous electron correlations with impurity scattering. The
scaling of $\rho_0$, $A$ and $n$ with pressure in Fig. 2 shows
that the same intrinsic physics is reflected in those parameters.

In a variety of systems, a regime of NFL behavior was ascribed to
strong fluctuations caused by the nearness of a quantum-critical
point, with the corollary that Fermi liquid behavior returns
gradually as the system is shifted away from the QCP, either with
changing pressure or composition, or with applying a magnetic
field to suppress spin fluctuations. In the case of BaVS$_3$, the only
obvious QCP \cite{asif} occurs when the critical value of the
control parameter $p_{\rm cr}$ is approached from the metallic
side, and the charge gap is suppressed. However, the extension of the
incommensurate magnetic order on the insulating side of the QCP
to its metallic side may have an influence on the NFL phenomenon.

The exponent $n=1.5$ is the canonical value expected at the QCP of
a nearly antiferromagnetic system \cite{ueda_moriya}. Samples
selected by the rigorous criteria described above do not include
any with $n<1.5$. These samples are the most pure BaVS$_3$
crystals, which suggests that the disorder is intrinsic. At the
border of the NFL regime $p=2.7$GPa the residual resistance of
$\rho_0=7.25\mu\Omega{\rm cm}$ corresponds to a mean free path of
130 V-V interatomic distances, while the system is known as
bad-metal with mean free path comparable to the lattice constant
at room temperature (or above). Another indication of the good
sample quality is that the Fermi liquid form of the
electron--electron scattering contribution $\Delta\rho=AT^2$ holds
up to approximately 15 K. The $A(p=2.7{\rm GPa})=0.14\mu\Omega{\rm
cm}/{\rm K}^2$ value on the Kadowaki--Woods plot would mean that
high-pressure BaVS$_3$ is analogous to the moderately heavy
fermion $d$-electron system with an expected $\gamma\sim
100$(mJ/mole K$^2$).

\begin{figure}[tb]
\centerline{\includegraphics[width=1\columnwidth] {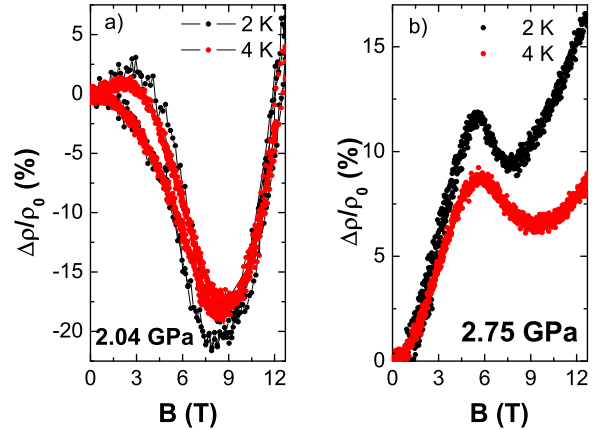}}
\caption{(Color online) Isothermal magnetoresistance curve
recorded at different temperatures and pressures. (a) The 2 and 4
K curves at 2.04 GPa. (b) Curves taken at 2.75 GPa for several
temperatures. The cyan arrow indicates the position of the dip for
various temperatures. } \label{fig3}
\end{figure}

Next we investigate a different aspect of the low-temperature
metallic phase. At ambient pressure BaVS$_3$ breaks, in
succession, point group, translational, and time reversal
symmetries, so the question emerges whether high-$p$ BaVS$_3$
would not undergo some kind of ordering transition. We attempted
to find the corresponding resistivity anomaly. At $p=2.03$GPa the
best candidate seems to be a weak shoulder around
$T_g\approx$15-20K, which separates the low-temperature NFL regime
from the higher-$T$ metallic regime in which $\rho$ rises
monotonically from 63 $\mu\Omega cm$ at 20K to $580 \mu\Omega cm$
at room temperature. This shoulder resembles (but is not as strong
as) the anomaly signaling the onset of ferromagnetic order in
BaVSe$_3$ \cite{bavse}. We also know from studies of the
near-critical regime on the insulating side that 15-20K is a
significant range of temperature where magnetic degrees of freedom
become observable \cite{Hysteresis}. This motivated extending our
magnetoresistivity studies to the high-pressure regime.

Figure 3a shows the magnetic field dependence of the
low-temperature resistivity at $p=2.04$GPa, i.e., slightly above
the critical pressure. The striking feature is the existence of
hysteresis which proves that metallic BaVS$_3$ possesses some kind
of magnetic order at these temperatures. As the only notable
feature of the temperature dependence of the resistivity is the
shoulder at $T_g$, we guess that this is the ordering temperature.
The nature of the order may be either structural distortion (such
as the low-pressure tetramerization) or magnetic order. We exclude
the former since the energetic motivation of tetramerization would
be the opening of spin and charge gaps. We also note that the MIT
has a structural \cite{Fagot} and resistivity \cite{Mihaly}
precursor and it is known that the resistivity precursor is
suppressed at the same pressure where the ground state becomes
metallic \cite{Forro}. Thus the $T<T_g$ high-pressure metallic
phase of BaVS$_3$ should be isostructural with the
ambient-pressure $T_{\rm MI}<T<T_S$ phase, and it is magnetically
ordered.

In our interpretation the quantum phase transition at $p=p_{\rm
cr}$ is a magnetic-insulator-to-magnetic-metal transition. The
high-pressure magnetic structure is unknown. It cannot be exactly
the same as at $p<p_{\rm cr}$, as it exists on a different
crystalline background, and spin--orbit coupling is relevant
\cite{Mihaly}. Still, it is likely to be similar to the
ambient-pressure low temperature magnetic state \cite{TxRef} in
the sense of having some long period \cite {albeit}. This allows
to interpret the $B$-cycle hysteresis shown in Figure 3a by
invoking our previous argument for similar phenomena in the
insulating phase \cite{Hysteresis}. We envisage that the effect of
changing the pressure and other thermodynamical parameters, like
$T$ and $B$, is changing {\bf Q}, the ordering vector of the
magnetic structure, or its aligning (casused by the relativistic
spin--orbit coupling which is manifest in the magnetic anisotropy
\cite{Mihaly}) with the lattice. Hysteresis should remain
observable as long as there is a magnetic order with finite
period. Since hysteresis is definitely absent at $p=2.75$GPa (Fig.
3b), where Fermi liquid behavior is restored, we suggest the
existence of a magnetic phase at $T<T_g$ in the $p-T$ domain with
unambiguously  NFL behavior.

NFL behavior is not typical of structurally ordered
metallic magnets. This shows that the magnetic order below the
$T_g(p)$ line is not necessarily the conventional long-range order
of either a SDW/AFM or a FM system. We rather propose a state
 analogous to the partially ordered extended NFL regime of MnSi
\cite{partial_MnSi}. It has been claimed that FL theory or its
straightforward extension fail to describe some ordered phases in
nearly/weakly FM materials \cite{nonFL_FM,FLbreakdown_MnSi}.
Lacking direct evidence, we would rather refrain from stating that
the similarity would go as far as high-pressure BaVS$_3$ being
actually a partially ordered FM \cite{albeit}. There may exist
generalizations of the novel metallic phase discussed for MnSi to
the systems with coupled magnetic and bond/charge orders.

In conclusion, clean high-quality BaVS$_3$ possesses an extended
NFL regime which at the same time has some kind of magnetic order.
The novel state of magnetism in good samples with mean free paths
$l\sim O(100{\AA})$ is characterized by diverging low-temperature
electron--electron scattering evident in $\Delta\rho\propto T^n$,
with exponent $1.5 \le n<2$, and hysteresis phenomena indicating
magnetic order.

Useful discussions with S. Bari\v{s}i\'{c}, J.P. Pouget, H.
Nakamura, T. Kobayashi, and I. Kup\v{c}i\'{c} are acknowledged.
This work was supported by the Swiss National Foundation for
Scientific Research and its research pool "MaNEP",  by the
Hungarian Research Funds OTKA TS049881, T049607 and Bolyai
00239/04 and by Croatian-Hungarian Intergovernmental S \& T
Cooperation Programme.


\begin{thebibliography}{1}
\bibitem{asif} The high pressure metal-insulator transition was not found within the experimental resolution to be discontinuous in pressure \cite{Hysteresis} so we assume that for the $T>2$K behavior at $p>p_{\rm cr}$ the $T=0$ transition can be regarded as a QCP.
\end{thebibliography}
\end{document}